\documentclass[conference,review]{llncs}
\institute{}
\usepackage{iftex}
\usepackage{upquote}
\usepackage[main=english]{babel}

\usepackage[hyphens]{url}

\makeatletter
\g@addto@macro{\UrlBreaks}{\UrlOrds}
\makeatother

\usepackage[%
    rm={oldstyle=false,proportional=true},%
    sf={oldstyle=false,proportional=true},%
    tt={oldstyle=false,proportional=true,variable=false},%
    qt=false%
]{cfr-lm}
\usepackage[T1]{fontenc}

\usepackage[
  babel=true, %
  expansion=alltext,
  protrusion=alltext-nott, %
  final %
]{microtype}

\DisableLigatures{encoding = T1, family = tt* }

\usepackage{diagbox}

\usepackage[dvipsnames, table]{xcolor}
\usepackage{fancyvrb}
\usepackage{newfloat}
\DeclareFloatingEnvironment[
  fileext = lol,          
  listname = {List of Listings}, 
  name = Listing,         
  placement = hbp         
]{listing}
\usepackage[labelfont=bf,font=small,skip=4pt,hypcap=true]{caption}
\SetupFloatingEnvironment{listing}{name=List.,within=none}

\usepackage[autostyle=true]{csquotes}

\defineshorthand{"`}{\openautoquote}
\defineshorthand{"'}{\closeautoquote}

\usepackage{xspace}
\makeatletter
\xspaceaddexceptions{\grqq \grq \csq@qclose@i \} }
\makeatother

\usepackage[%
  square,        %
  comma,         %
  numbers,       %
  sort           %
]{natbib}

\usepackage{etoolbox}
\makeatletter
\patchcmd{\NAT@test}{\else \NAT@nm}{\else \NAT@hyper@{\NAT@nm}}{}{}
\makeatother

\SetExpansion
[ context = sloppy,
  stretch = 30,
  shrink = 60,
  step = 5 ]
{ encoding = {OT1,T1,TS1} }
{ }

\usepackage[rflt]{floatflt}

\usepackage{multirow} %

\usepackage{dcolumn}  %

\usepackage{stfloats}
\fnbelowfloat

\usepackage{pdfcomment}


\usepackage[caption=false,font=footnotesize]{subfig}

\usepackage{mindflow}

\DeclareFontFamily{U}{MnSymbolC}{}
\DeclareSymbolFont{MnSyC}{U}{MnSymbolC}{m}{n}
\DeclareFontShape{U}{MnSymbolC}{m}{n}{
  <-6>    MnSymbolC5
  <6-7>   MnSymbolC6
  <7-8>   MnSymbolC7
  <8-9>   MnSymbolC8
  <9-10>  MnSymbolC9
  <10-12> MnSymbolC10
  <12->   MnSymbolC12%
}{}
\DeclareMathSymbol{\powerset}{\mathord}{MnSyC}{180}

\ifpdftex
\input glyphtounicode
\fi

\newenvironment{tightcenter}{%
  \setlength\topsep{0pt}
  \setlength\parskip{0pt}
  \begin{center}
}{%
  \end{center}
}

\usepackage{color,tikz}
\usepackage{wrapfig}
 \usepackage{pict2e}
\usepackage{graphicx} %
\usepackage{wrapfig}
\usepackage{todonotes}
  \newcounter{todocounter}

\usetikzlibrary{external}
\usepackage{listings}
\usepackage{comment}
\usepackage{booktabs}
\newcommand*{\otoprule}{\midrule[\heavyrulewidth]}
\usepackage{multirow}
\usepackage{makecell}
\usepackage{amsfonts,amssymb,amsmath}

\usepackage[capitalise,nameinlink]{cleveref}

\crefname{section}{Sect.}{Sect.}
\Crefname{section}{Section}{Sections}
\crefname{listing}{List.}{List.}
\crefname{listing}{Listing}{Listings}
\Crefname{listing}{Listing}{Listings}

\usepackage{mathtools}

\usepackage{xcolor}
\usepackage{colortbl}

\definecolor{myGold}{RGB}{230,159,0} %
\definecolor{mySkyBlue}{RGB}{86,180,233} %
\definecolor{myGreen}{RGB}{0,158,115} %
\definecolor{myYellow}{RGB}{240,228,66} %
\definecolor{myBlue}{RGB}{0,114,178} %
\definecolor{myRed}{RGB}{213,94,0} %
\definecolor{myPink}{RGB}{204,121,167} %

\usepackage{tcolorbox}

\usepackage{graphicx}
\usepackage[retainorgcmds]{IEEEtrantools}
\usepackage{pifont}
\newcommand*{\cmark}{\ding{51}}
\newcommand*{\xmark}{\ding{55}}
\usepackage{ltl}

\makeatletter
\DeclareRobustCommand{\onontimes}{%
  \mathbin{\mathpalette\on@ntimes\relax}%
}
\newcommand{\on@ntimes}[2]{%
  \vcenter{\hbox{%
    \sbox0{\m@th$#1\otimes$}%
    \setlength\unitlength{\wd0}%
    \begin{picture}(1,1)
    \linethickness{0.35pt}
    \put(.5,.5){\circle{.85}}
    \end{picture}%
  }}%
}
\makeatother
\usepackage{paralist}

\usetikzlibrary{automata, arrows.meta, positioning, shapes}

\tikzset{
	state/.style=
    {circle, draw, align=center, auto, initial text={}}, 
	>=stealth,
	loopright/.style={loop,looseness=5,out=35, in=-35},
	loopleft/.style={loop,looseness=5,out=145, in=215},
	loopabove/.style={loop,looseness=5,out=125, in=55},
	loopbelow/.style={loop,looseness=5,out=-125, in=-55}
}

\usetikzlibrary{fit} %
\usetikzlibrary{backgrounds} %
\usetikzlibrary{patterns}
\usetikzlibrary{shapes}
\usetikzlibrary{snakes}
\usetikzlibrary{shadows}
\usetikzlibrary{arrows}
\tikzstyle{cirre}=[draw=green!60!red, fill=green!20!white,circle,minimum size=1.4em,inner sep=0em]                                                                                                      
\tikzstyle{cir1re}=[draw=red!80!violet, fill=red!20!white,circle,minimum size=1.4em,inner sep=0em]  

\tikzstyle{cir}=[draw=violet, fill=red!20!white,circle,minimum size=1.4em,inner sep=0em]                                                                                                      
 \tikzstyle{dia}=[draw=green!80!red, fill=green!20!white, diamond,minimum size=1.4em,inner sep=0.1em]                                                                                                      
\tikzstyle{cir1}=[draw=violet, fill=violet!20!white,circle,minimum size=1.4em,inner sep=0em]

\tikzstyle{background}=[rectangle,fill=gray!10, inner sep=0.1cm, rounded corners=0mm]
\tikzstyle{loc}=[draw,rectangle,minimum size=1.4em,inner sep=0em]
\tikzstyle{trans}=[-latex, rounded corners]
\tikzstyle{trans2}=[-latex, dashed, rounded corners]
\tikzset{snake it/.style={decorate, decoration=snake}}%

\tikzstyle{inv}=[]
\tikzstyle{varpass}=[]

\newcommand{\until}{\mathbin{\mathbf{{U}}}}
\newcommand{\release}{\mathbin{\mathbf{{R}}}}
\newcommand{\since}{\mathbin{\mathbf{{S}}}}
\newcommand{\trigger}{\mathbin{\mathbf{{T}}}}

\newcommand{\fmtl}{\textmd{\textup{\textsf{Flat-MTL}}}}
\newcommand{\stl}{\textmd{\textup{\textsf{STL}}}}
\newcommand{\mtl}{\textmd{\textup{\textsf{MTL}}}}
\newcommand{\ltl}{\textmd{\textup{\textsf{LTL}}}}
\newcommand{\mitlpp}{\textmd{\textup{\textsf{MITPPL}}}}

\newcommand{\mightyppl}{\textmd{\textup{\textsc{MightyPPL}}}}
\newcommand{\mightyl}{\textmd{\textup{\textsc{MightyL}}}}
\newcommand{\tempora}{\textmd{\textup{\textsc{Tempora}}}}
\newcommand{\mitl}{\textmd{\textup{\textsf{MITL}}}}
\newcommand{\mtlpp}{\textmd{\textup{\textsf{MTLPPL}}}}

\newcommand*{\ta}[1]{\mathsf{TA}\text{#1}}
\newcommand*{\dc}[1]{} %

\newcommand*{\AP}{\mathsf{AP}}

\newcommand{\diamondminus}{%
  \sbox0{$\lozenge$}%
  \usebox0\kern-.5\wd0\clap{\raisebox{.1ex}{\scalebox{.7}[1]{$-$}}}\kern.5\wd0%
}
\DeclareMathOperator{\past}{\once}

\DeclareMathOperator{\nm}{\mathbf{Y}}
\DeclareMathOperator{\nex}{\mathbf{X}}

\newcommand{\R}{\mathbb{R}_{\geq 0}}
\newcommand{\N}{\mathbb{N}}

\newcommand{\I}{\mathbb{I}}

\newcommand*\sem[1]{\ensuremath{\llbracket#1\rrbracket}}

\definecolor{saffron}{rgb}{1.0,0.49,0.0}

\usepackage{microtype}
\makeatletter
\g@addto@macro\@verbatim{\microtypesetup{activate=false}}
\makeatother

\makeatletter
\newcommand{\oset}[3][0ex]{%
  \mathrel{\mathop{#3}\limits^{
    \vbox to#1{\kern-2\ex@
    \hbox{$\scriptstyle#2$}\vss}}}}
\makeatother

\newcommand{\Pnkern}{%
  \mkern-2mu
}

\DeclareMathOperator{\eventually}{\mathbf{F}}

\DeclareMathOperator{\once}{\overset{\leftarrow}{\mathbf{F}}}
\DeclareMathOperator{\globally}{\mathbf{G}}
\DeclareMathOperator{\henceforth}{\overset{\leftarrow}{\mathbf{G}}}
\DeclareMathOperator{\nextx}{\mathbf{X}}

\DeclareMathOperator{\PnF}{\mathbf{P \Pnkern n}}
\DeclareMathOperator{\PnO}{\oset[-1pt]{\leftarrow}{\mathbf{P \Pnkern n}}}
\DeclareMathOperator{\dualPnF}{\vphantom{\mathbf{P \Pnkern n}}\smash{{\mathbf{P \Pnkern n}}^{\mathrlap{\sim}}}}
\DeclareMathOperator{\dualPnO}{\vphantom{\mathbf{P \Pnkern n}}\smash{\oset[-1pt]{\leftarrow}{\mathbf{P \Pnkern n}}}^{\mathrlap{\sim}}}

\makeatletter
\renewcommand{\paragraph}{\@startsection{paragraph}{6}{\z@}{2ex}{-0.7em}{\normalsize\bf}}
\makeatother

\usepackage{algpseudocode}
\usepackage{algorithm}

\usepackage{stmaryrd}
\usepackage{paralist}

\usepackage{hyperref}

\hypersetup{
  colorlinks=true,       %
  raiselinks=true,       %
  pdfstartview=Fit,
  breaklinks=true,       %
  hypertexnames=false,   %
}

\title{$\mightyppl{}$: Towards Model Checking \mtl{}}
\titlerunning{\hyperlink{page:toc}{$\mightyppl{}$: Towards Model Checking \mtl{}}}

 \author{%
     Hsi-Ming Ho\inst{1} \and
     Shankara Narayanan Krishna\inst{2} \and
     Khushraj Madnani\inst{4} \and
     Rupak Majumdar\inst{3} \and
     Paritosh Pandya\inst{2}
 }

  \institute{
      University of Sussex, Brighton, United Kingdom\\
      \and
      Indian Institute of Technology Bombay, Mumbai, India\\
      \and
      Max Planck Institute for Software Systems (MPI-SWS), Kaiserslautern, Germany\\
      \and
      Indian Institute of Technology Guwahati, Guwahati, India\\
}

\begin{document}

\maketitle

\begin{abstract}

The theoretical foundation for model checking timed systems against
\emph{Metric Interval Temporal Logic} ($\mitl{}$) was established in the early 1990s,
yet the first practical tool supporting future $\mitl{}$ ($\mightyl{}$) did not emerge until 2017.
Recently, there has been growing interest in extending this toolchain to support more expressive logical operators, including \emph{past modalities}, \emph{Pnueli modalities}, and limited use of \emph{singular intervals}. \mightyppl{} is one such toolchain. 
We introduce an upgraded version of  $\mightyppl{}$ that
enables for the first time, the model checking of \emph{Metric Temporal Logic} ($\mtl{}$)
properties of the form $\globally (p \Rightarrow \eventually_{=5} q)$ in addition to Pnueli and Past modalities. 
We discuss the tool’s
underlying architecture and implementation, and present a performance evaluation
against the \textsc{Tempora} tool across diverse satisfiability and model checking
benchmarks, demonstrating that $\mightyppl{}$ delivers significantly better
performance.

\end{abstract}

\section{Introduction}
\label{sec:intro}

\paragraph{Timed logics and automata.}
\emph{Metric Temporal Logic} ($\mtl{}$)~\cite{koymans} extends \emph{Linear Temporal Logic} ($\ltl{}$)~\cite{4567924}
with temporal operators decorated by \emph{time intervals}; e.g.,~`each request is followed by an
acknowledgement after \emph{exactly $5$ seconds}' is written
$\globally (\textit{req} \Rightarrow \eventually_{[5, 5]} \textit{ack})$.
$\mtl{}$ and its signal variant $\stl{}$~\citep{maler2004monitoring} are widely used in
cyber-physical systems.
\emph{Timed automata} ($\ta{s}$)~\cite{AD94} are the \emph{de facto} model for timed systems and
benefit from decades of mature zone-based tooling (\textsc{Uppaal}~\cite{behrmann2007uppaal},
\textsc{LTSmin}~\cite{laarman2013multi}, \textsc{TChecker}~\cite{TChecker}).
Solving satisfiability and model checking for $\mtl{}$ therefore reduces, in principle, to
translating formulae into $\ta{s}$.
Such translations are, however, only possible when the logic is restricted to
\emph{Metric Interval Temporal Logic} ($\mitl{}$), i.e., $\mtl{}$ without \emph{singular (`punctual')
intervals}~\cite{AFH96}.
Until recently, \textsc{MightyL}~\cite{DBLP:conf/cav/BrihayeGHM17} was the only tool translating
future $\mitl{}$ formulae into $\ta{s}$; its successor $\mightyppl{}$~\cite{mightypplpaper}
added \emph{past modalities} and \emph{Pnueli modalities} but still disallowed punctual intervals.

\begin{figure}[!htbp]
    \centering
    \includegraphics[width=0.75\linewidth]{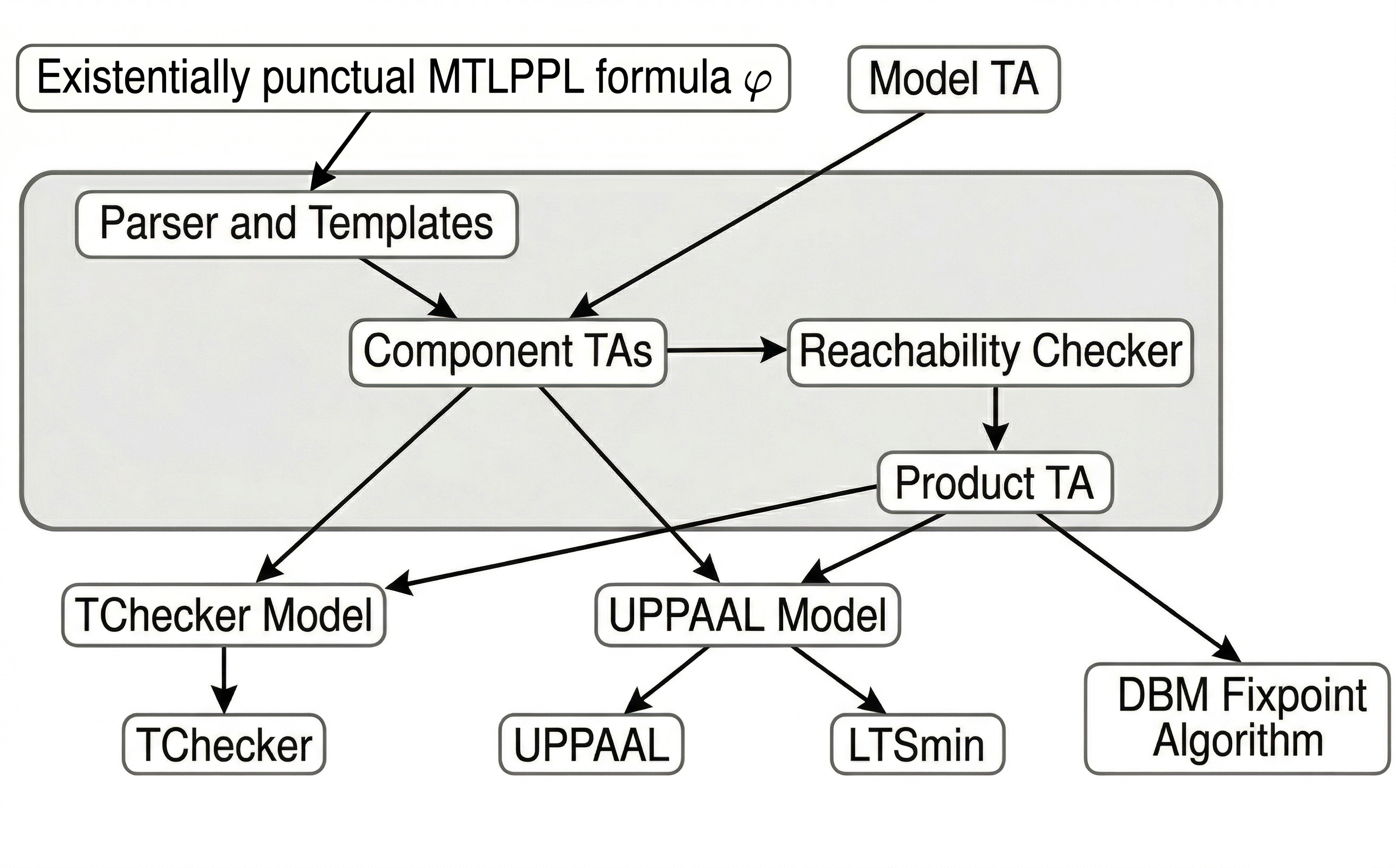}
    \captionof{figure}{$\mightyppl{}$ architecture.}
    \label{fig:system}
\end{figure}

\paragraph{Contributions.}
We present an updated version of $\mightyppl{}$ that, for the first time in a practical tool,
handles a useful class of formulae with punctual intervals including
`one-sided' $\mtl{}$~\cite{bouyer2025model} with past and Pnueli modalities~\cite{mightypplpaper}.
Our contributions, relative to~\cite{mightypplpaper} and to prior tools, are:
(i) an end-to-end implementation with multiple interchangeable backends
(\textsc{Uppaal}, \textsc{LTSmin}, \textsc{TChecker}, and a built-in symbolic fixpoint algorithm);
(ii) a reachability-driven product construction (flattening) that
keeps monolithic $\ta{s}$ small enough for explicit analysis; and
(iii) an extensive empirical comparison against \textsc{Tempora}~\cite{temporapaper}, the only other tool
supporting $\mitl{}$ with past modalities, on satisfiability and model-checking benchmarks.
On the benchmarks in \cref{sec:exp}, $\mightyppl{}$ outperforms \textsc{Tempora} by
orders of magnitude on the majority of instances, scales to multi-core model checking via
\textsc{LTSmin}, and produces correct answers on all instances whereas \textsc{Tempora} 
reports wrong results on several instances.

\paragraph{Related work.}
A fragment of $\mtl{}$ 
with a decidable satisfiability problem, namely,  $\fmtl{}$, which subsumes the `one-sided' $\mtl{}$ of~\cite{bouyer2025model} was studied in~\cite{Bouyer2007,Bouyer2008}. 
To the best of our knowledge, however, no tool supporting $\fmtl{}$ has been implemented.
 Beyond $\mightyl{}$ and the original $\mightyppl{}$ of~\cite{mightypplpaper},
\textsc{Tempora} is the most closely related tool; it follows the
spirit of~\cite{maler2006mitl} but is formulated in the framework of
\emph{generalized timed automata}~\cite{akshay2023unified}.
\textsc{Tempora} allows punctual intervals, but in a highly restricted manner: \emph{they are permitted only at the outermost level of a formula}.
None of these support past together with Pnueli modalities. 

\begin{table}[!htbp]
\caption{Feature coverage of $\mightyppl{}$ vs.\ prior tools.
\cmark = supported, \xmark = not supported, (\cmark) = partially supported.}
\label{tab:feature-main}
\centering
\scalebox{0.85}{
\begin{tabular}{lccc}
\toprule
Feature & \textsc{MightyL} \hspace*{1em} & \textsc{Tempora} \hspace*{1em} & $\mightyppl{}$ \\
\midrule
Future $\mitl{}$                        & \cmark   & \cmark   & \cmark  \\
Past modalities                         & \xmark   & \cmark   & \cmark  \\
Pnueli modalities                       & \xmark   & \xmark   & \cmark  \\
Punctual intervals (one-sided $\mtl{}$) & \xmark   & (\cmark) & \cmark  \\
Symbolic alphabets                      & \xmark   & \xmark   & \cmark  \\
Multiple backend support                & (\cmark) & \xmark   & \cmark  \\
Built-in fixpoint algorithm             & \xmark   & \xmark   & \cmark  \\
\bottomrule
\end{tabular}}
\end{table}

\section{Preliminaries}
\label{sec:prelim}
We refer the reader to~\cite{mightypplpaper} for more detailed definitions.

\paragraph{Timed words.} Let $\R, \N$ denote the non-negative reals and naturals; let $\I$ (resp.\ $\I_0$) denote
intervals $\langle l, u\rangle$ with $l\le u\in\N\cup\{\infty\}$ (resp.\ with $l=0$).
Fix a finite set $\AP$ of atomic propositions and let $\Sigma_\AP = 2^\AP$.
An infinite (resp.~finite) \emph{timed word} over $\Sigma_\AP$ is an infinite (resp.~finite) non-Zeno (weakly) monotonic sequence
$\rho = (\sigma_1,\tau_1)(\sigma_2,\tau_2)\dots$ with $\sigma_i\in\Sigma_\AP$ and $\tau_i\in\R$.

\paragraph{$\mtl{}$ with Past and Pnueli modalities ($\mtlpp$)~\cite{mightypplpaper}.} $\mtlpp$
formulae over $\AP$ are defined inductively as 
\[
\varphi ::= p \mid \neg\varphi \mid \varphi\wedge\varphi \mid \varphi\until_I\varphi
          \mid \varphi\since_I\varphi \mid \PnF_J(\varphi_1,\dots,\varphi_k)
          \mid \PnO_J(\varphi_1,\dots,\varphi_k),
\]
where $a\in\AP$, $I\in\I$, $J\in\I_0$.
We also write psuedo-arithmetic expressions for intervals, e.g., `$=5$' for $[5, 5]$.
The pointwise semantics is standard on atomic propositions, conjunctions and negations; $\until_I$ and $\since_I$ are the (strict) future and
past metric versions of `Until' and `Since' of $\ltl{}$.
The Pnueli modalities~$\PnF_J$ and $\PnO_J$ assert the existence of $n$ strictly increasing
(resp.~decreasing) positions $i_n > \dots > i_1 > i$ with $\tau_{i_k}-\tau_i \in J$ and
$\rho,i_k\models \varphi_k$.
We also define the duals of these modalities: $\varphi_1 \release_I \varphi_2 \equiv \neg ((\neg
\varphi_1) \until_I (\neg \varphi_2))$,  $\varphi_1 \trigger_I \varphi_2 \equiv \neg ((\neg
\varphi_1) \since_I (\neg \varphi_2))$, $\dualPnF_I(\varphi_1, \dots, \varphi_n) \equiv \neg \PnF_I(\neg
\varphi_1, \dots, \neg \varphi_n)$, $\dualPnO_I(\varphi_1, \dots, \varphi_n) \equiv \neg \PnO_I(\neg
\varphi_1, \dots, \neg \varphi_n)$.
Other derived operators like $\eventually_I,\once_I,\globally_I,\henceforth_I,\nex,\nm$ are defined 
as usual.  The \emph{timed language} of $\varphi$ is defined as $\sem{\varphi} = \{\rho | \rho, 1 \models \varphi\}$.

\paragraph{Existentially punctual $\mtlpp$.}
An $\mtlpp{}$ formula is \emph{existentially punctual} if, when written in negation normal form,
all of its punctual subformulae only appear in
\begin{inparaenum}[(i)]
\item $\varphi_2$ in larger subformulae $\varphi_1 \until_I \varphi_2$ or $\varphi_1 \since_I \varphi_2$;
\item $\varphi_1$ in larger $\varphi_1 \release_I \varphi_2$ 
    or $\varphi_1 \trigger_I \varphi_2$;
\item $\varphi_k$ ($1 \leq k \leq n$) in larger $\PnF_I(\varphi_1, \dots, \varphi_n)$ or $\PnO_I(\varphi_1, \dots, \varphi_n)$.
\end{inparaenum}
 For example, $\eventually(p \land \eventually_{=5} q)$ and 
$\neg \globally_{[0, 20]} (\textit{req} \Rightarrow ((\neg \once_{[0, 2]} (\textit{ack}_2 \land \once_{=1} \textit{ack}_1)) \until_{=10} \textit{grant} ))$
are existentially punctual.
`One-sided' $\mtl{}$ of~\cite{bouyer2025model} is the fragment of existentially punctual $\mtlpp{}$ without past and Pnueli modalities.

\paragraph{Timed automata and model checking.}
A \emph{timed automaton} ($\ta{}$) $\mathcal{A}$ over $\Sigma_\AP$ is a finite
automaton equipped with real-valued clocks, clock constraints on transitions, and
clock resets~\cite{AD94}; $\mathcal{A}$ accepts a timed language $\sem{\mathcal{A}}$ under
B\"uchi or finite-word acceptance.
Given a formula $\varphi \in \mtlpp$ and a model $\ta{}$ $\mathcal{M}$, the
\emph{satisfiability} problem asks whether $\sem{\varphi} \neq \emptyset$, and the
\emph{model-checking} problem asks whether
$\sem{\mathcal{M}} \subseteq \sem{\varphi}$, i.e., whether every behaviour of $\mathcal{M}$
satisfies $\varphi$.

\section{Tool Design and Implementation}
\label{sec:imp}

\paragraph{Overview.} The tool $\mightyppl{}$ is a command-line program written in C\texttt{++}17.
Given an existentially punctual $\mtlpp{}$ formula, $\mightyppl{}$ can output component $\ta{s}$ individually (the `\texttt{compflat}' mode), or generate the synchronous product of component $\ta{s}$
as a single monolithic $\ta{}$ (the `\texttt{flat}' mode).
The system is depicted in~\cref{fig:system}.
The source code repository is hosted on GitHub at
\begin{tightcenter}
\url{https://github.com/hsimho/MightyPPL}
\end{tightcenter}
\paragraph{Symbolic encoding of atomic propositions.}
In $\mightyl{}$ and $\textsc{Tempora}$, atomic propositions are represented by Boolean variables which are set to $\textbf{true}$ or $\textbf{false}$ non-deterministically.
$\mightyppl{}$ works similarly in the `\texttt{compflat}' mode,
but it also uses wildcard `$\ast$' values to effectively force the back-end tools to perform `symbolic execution' (rather than `exhaustive testing').

\paragraph{Forward and backward reachability analysis.}
$\mightyppl{}$ uses BDDs to synchronise transitions of component $\ta{s}$ into joint transitions in the `\texttt{flat}' mode.
It enumerates only the combinations of
component transitions whose symbolic label conjunctions are satisfiable, and generates only
locations and transitions forward-reachable from the initial product location; a backward
reachability pass further rules out locations with no accepting run.

\paragraph{Using the tool.}
$\mightyppl{}$ is invoked as
\begin{tightcenter}
\verb/mitppl <in_spec_file> --{fin|inf} [out_file --{tck|xml}]/
\end{tightcenter}
where \verb/in_spec_file/ contains the $\mitlpp{}$ formula and \texttt{fin}/\texttt{inf}
selects finite-/infinite-word semantics; the optional output file format is
\texttt{tck}/\texttt{xml}.
The flag \texttt{--compflatten} 
selects the `\texttt{compflat}' mode;
For model checking, the model $\mathcal{M}$ can be specified by editing
\texttt{MightyPPL.cpp} or the generated files.

\section{Experiments}
\label{sec:exp}

All experiments were conducted on a desktop machine with an Intel i9-13900K CPU and 64GB memory. We used commit \href{https://github.com/ticktac-project/tchecker/tree/d711ace9ff754d8d952f5d491a31591115300c7f}{\texttt{d711ace}} of \textsc{TChecker},
commit \href{https://github.com/DEIS-Tools/MoniTAal/tree/4bbf3cc1b199c881e99e0d7f71dffefd969c13a2}{\texttt{4bbf3cc}} of \textsc{MoniTAal},
commit \href{https://github.com/DEIS-Tools/PARDIBAAL/tree/1c8f7c2f64cd2febe0d7874862c768d627998d31}{\texttt{1c8f7c2}} of PARDIBAAL, 
commit \href{https://github.com/ssoelvsten/buddy/tree/5aca063a4b2e90352480f3dd24daeb6dbefa2d33}{\texttt{5aca063}} of BuDDy.
The main baseline is \textsc{Tempora}~\cite{temporapaper} (commit \href{https://github.com/EQuaVe/TEMPORA/tree/1ba0bad25a9015a077f5990b93954bc2157f0186}{\texttt{1ba0bad}}), the only other tool supporting $\mitl{}$ with past modalities.

\paragraph{Simple $\mtl{}$ formulae.}

We consider the finite- and infinite-word satisfiability of the handcrafted $\mtl{}$ formulae below:
{\small \allowdisplaybreaks
\begin{IEEEeqnarray*}{rClrCl}
\varphi_1 &=& \eventually \big( (p \until_{[2, 3]} q) \land \neg (p \until_{[0, 3]} q) \big) \;, 
&\varphi_2 &=& \eventually \big( (p \since_{[2, 3]} q) \land \neg (p \since_{[0, 3]} q) \big) \;, \\
\varphi_3 &=& \globally ( \neg p \lor \eventually_{[1, \infty)} q) \land \eventually_{[2, 3]} \big( p \land \neg (\eventually q) \big) \;, 
& \varphi_4 &=& \eventually_{[5, \infty)} \big( p \land (\top \since_{[2, 3]} q) \big) \land \globally (\neg q) \;,  \\
\varphi_5 &=& (\eventually_{[1, 2]} p) \land \globally_{[1,3]}\big( \neg p \lor (\top \since_{[1,2]} q)\big) \;, 
& \varphi_6 &=& (\eventually_{[1, 2]} p) \land \globally_{[1,3]}\big( \neg p \lor (\top \since_{[3,6]} q)\big) \;,  \\
\varphi_7 &=& \eventually \Big( \big( p \since_{[1, 3]} (\eventually_{[2,5]} q) \big) \land \globally (\neg q) \Big) \;, 
& \varphi_8 &=& \eventually \Big( \big( p \since_{[1, 3]} (\eventually_{=2} q) \big) \land \globally (\neg q) \Big) \;,  \\
\varphi_9 &=& \eventually \Big( \big( p \since_{[1, 3]} (\eventually_{=4} q) \big) \land \globally (\neg q) \Big) \;, 
& \varphi_{10} &=& \eventually \Big( \big( p \since_{[1, 5]} (\eventually_{[2,3]} q) \big) \land \neg (p \since q) \Big) \;,  \\
\varphi_{11} &=& \eventually \Big( \big( p \since_{=2} (\eventually_{[2,3]} q) \big) \land \neg (p \since q) \Big) \;, 
& \varphi_{12} &=& \eventually \Big( \big( p \since_{=4} (\eventually_{[2,3]} q) \big) \land \neg (p \since q) \Big) \;,  \\
\varphi_{13} &=& \eventually \Big( \big( \eventually_{[2,5]} (p \since_{[1,3]} q) \big) \land \globally (\neg q) \Big) \;, 
& \varphi_{14} &=& \eventually \Big( \big( \eventually_{[2,5]} (p \since_{=3} q) \big) \land \globally (\neg q) \Big) \;, \\
\varphi_{15} &=& \eventually \Big( \big( \eventually_{[2,5]} (p \since_{=1} q) \big) \land \globally (\neg q) \Big) \;, 
& \varphi_{16} &=& \eventually \Big( \big( \eventually_{[2,3]} (p \since_{[1,5]} q) \big) \land \neg (p \since q) \Big) \;,  \\
\varphi_{17} &=& \eventually \Big( \big( \eventually_{=3} (p \since_{[2,5]} q) \big) \land \neg (p \since q) \Big) \;, 
& \varphi_{18} &=& \eventually \Big( \big( \eventually_{=1} (p \since_{[2,5]} q) \big) \land \neg (p \since q) \Big) \;, 
\end{IEEEeqnarray*}
}%
\vspace*{-1.5em}
{\allowdisplaybreaks \small
\begin{IEEEeqnarray*}{rCl}
\varphi_{19} &=& \eventually \bigg( (p \since_{[1,2]} q) \land (p \since_{[3,4]} r) \land \neg\Big(\top \since \big(q \land (p \since_{[1,2]} r)\big)\Big) \bigg) \;, \\
\varphi_{20} &=& \eventually \bigg( (p \since_{[1,2]} q) \land (p \since_{[3,4]} r) \land \neg\Big(\top \since \big(q \land (p \since_{[1,3]} r)\big)\Big) \bigg) \;,  \\
\varphi'_{i} &=& \eventually\big( \globally_{[0, 5]}(p_1 \land \dots \land p_i) \land \eventually_{[0, 3]} (\neg p_i) \big) 
\quad \text{for } 10 \leq i \leq 15.
\end{IEEEeqnarray*}
}%
The results are listed in~\cref{tab:simplemitl}.
It is clear that satisfiability in the infinite-word setting is more challenging
for both tools.
$\tempora{}$ performs better on some testcases (e.g.,~$\varphi_5, \varphi_6$) but much worse on others (e.g.,~$\varphi_2, \varphi_{19}, \varphi_{20}$).
It is also notable that
$\tempora{}$ outputs incorrect results on some testcases ($\varphi_7, \varphi_{10}, \varphi_{13}, \varphi_{16}$), 
The results on $\varphi'_{10}, \dots \varphi'_{15}$ indicate that the runtime of \textsc{Tempora} can be \emph{exponential} in the number of atomic propositions, even when the formulae are intuitively simple, while $\mightyppl{}$ does not suffer from the same problem.

\begin{table}[!htbp]                    
\captionof{table}{Execution times on the simple $\mtl{}$ benchmarks.
Times are in seconds.
`TO' indicates timeouts (600s).
`-' means unsupported.
Incorrect results are marked by `*'.
} %
\label{tab:simplemitl}
\centering
\scalebox{0.8}{
\def\arraystretch{1.0}
\setlength\tabcolsep{1mm}
\begin{tabular}{l@{\hspace{2mm}}r>{\columncolor{mySkyBlue!10}}r>{\columncolor{myYellow!10}}r>{\columncolor{myBlue!45}}r>{\columncolor{myYellow!45}}r>{\columncolor{myBlue!45}}r>{\columncolor{myYellow!45}}r}
\toprule %
&
 & \cellcolor{white}\multirowcell{3}{$\tempora{}$ \\ \small \texttt{infinite}\\{}}
& \cellcolor{white}\multirowcell{3}{$\tempora{}$ \\ \small \texttt{finite}\\{}}
 & \cellcolor{white}\multirowcell{3}{$\mightyppl{}$ \\ \small \texttt{tck -{}-inf} \\ \small \texttt{compflat}}
 & \cellcolor{white}\multirowcell{3}{$\mightyppl{}$ \\ \small \texttt{tck -{}-fin} \\ \small \texttt{compflat}} 
 & \cellcolor{white}\multirowcell{3}{$\mightyppl{}$ \\ \small \texttt{tck -{}-inf} \\ \small \texttt{flat}}
 & \cellcolor{white}\multirowcell{3}{$\mightyppl{}$ \\ \small \texttt{tck -{}-fin} \\ \small \texttt{flat}} \\ \\
$\varphi$ & Sat? \\
\otoprule
$\varphi_{1}$            & \xmark            & 3.675    & 0.013    & 0.008         & 0.008     & 0.000      & 0.000      \\
$\varphi_{2}$            & \xmark            & TO       & 0.010    & 0.013         & 0.009     & 0.000      & 0.000     \\
$\varphi_{3}$            & \xmark            & 0.061    & 0.005    & 0.034         & 0.042     & 0.031      & 0.013         \\
$\varphi_{4}$            & \xmark            & 0.000    & 0.061    & 0.013         & 0.009     & 0.000      & 0.000     \\
$\varphi_{5}$            & \cmark            & 0.004    & 0.004    & 0.084         & 0.088     & 22.410     & 2.829         \\
$\varphi_{6}$            & \xmark            & 0.021    & 0.005    & 0.185         & 0.094     & 22.038     & 2.934     \\
$\varphi_{7}$            & \cmark            & TO       & 0.191*   & 0.008         & 0.009     & 0.013      & 0.008     \\ 
$\varphi_{8}$            & \cmark            & -        & -        & 0.008         & 0.008     & 0.012      & 0.008         \\
$\varphi_{9}$            & \xmark            & -        & -        & 0.009         & 0.010     & 0.013      & 0.008         \\  
$\varphi_{10}$           & \cmark            & 201.135* & 0.049*   & 0.009         & 0.008     & 0.022      & 0.012     \\ 
$\varphi_{11}$           & \cmark            & -        & -        & 0.009         & 0.009     & 0.021      & 0.012        \\
$\varphi_{12}$           & \xmark            & -        & -        & 0.009         & 0.012     & 0.022      & 0.012        \\
$\varphi_{13}$           & \cmark            & TO       & 0.203*   & 0.007         & 0.010     & 0.012      & 0.008  \\
$\varphi_{14}$           & \cmark            & -        & -        & 0.009         & 0.007     & 0.011      & 0.008        \\
$\varphi_{15}$           & \xmark            & -        & -        & 0.010         & 0.011     & 0.012      & 0.008        \\
$\varphi_{16}$           & \cmark            & 113.864* & 0.055*   & 0.008         & 0.010     & 0.016      & 0.011     \\
$\varphi_{17}$           & \cmark            & -        & -        & 0.010         & 0.008     & 0.017      & 0.010         \\
$\varphi_{18}$           & \xmark            & -        & -        & 0.011         & 0.012     & 0.017      & 0.010         \\
$\varphi_{19}$           & \cmark            & TO       & 0.120    & 23.002        & 0.787     & TO         & 231.522      \\
$\varphi_{20}$           & \xmark            & TO       & 3.520    & 565.107       & 4.187     & TO         & 227.865 \\ 
$\varphi'_{10}$          & \xmark            & 0.494    & 0.020    & 0.008         & 0.008     & 0.010      & 0.008    \\ 
$\varphi'_{11}$          & \xmark            & 1.230    & 0.038    & 0.008         & 0.008     & 0.009      & 0.008    \\ 
$\varphi'_{12}$          & \xmark            & 3.115    & 0.076    & 0.008         & 0.008     & 0.009      & 0.007    \\ 
$\varphi'_{13}$          & \xmark            & 9.274    & 0.191    & 0.008         & 0.008     & 0.009      & 0.007    \\ 
$\varphi'_{14}$          & \xmark            & 29.443   & 0.546    & 0.007         & 0.007     & 0.010      & 0.008    \\ 
$\varphi'_{15}$          & \xmark            & TO       & 1.983    & 0.007         & 0.007     & 0.010      & 0.008    \\ 
\bottomrule     
\end{tabular}     
}     
\end{table}

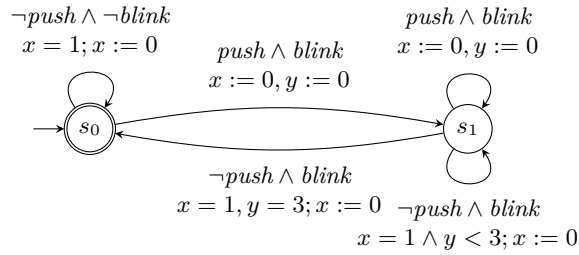
\begin{figure}
\centering
\begin{tikzpicture}[->, node distance=5cm, transform shape, scale=1]
   \node[initial left,state, accepting](0){$s_0$};
   \node[state, right of=0](1){$s_1$};
   
   \path
   (0) edge[loopabove] node[above=1mm, align=center, looseness=20, out=120, in=90]{$\neg \textit{push} \land \neg \textit{blink}$ \\ $x = 1; x := 0$} (0)
   (0) edge[->, bend left=10] node[above=1mm, align=center]{$\textit{push} \land \textit{blink}$ \\ $x := 0, y := 0$} (1)
   (1) edge[loopabove] node[above=1mm,align=center]{$\textit{push} \land \textit{blink}$ \\ $x := 0, y := 0$} (1)
   (1) edge[loopbelow] node[below=1mm, align=center]{$\neg \textit{push} \land \textit{blink}$ \\ $x = 1 \land y < 3; x := 0$} (1)
   (1) edge[->, bend left=10] node[below=1mm, align=center]{$\neg \textit{push} \land \textit{blink}$ \\ $x = 1, y = 3; x := 0$} (0);
 \end{tikzpicture}
\captionof{figure}{The model $\ta{}$ $\mathcal{M}_\textit{lamp}$.}
\label{fig:timedlamp}
\end{figure}

\paragraph{Timed lamp.}
This is a simple case study adapted from~\citep{bersani2016tool}.
We consider a lamp with a single button controlled by the user. Whenever the user \emph{pushes} the button, the lamp \emph{blinks} at the same instant and enters a state, where it blinks at exactly every $1$ time unit for three more times. The simple system can be modelled by the $\ta{}$ $\mathcal{M}_\textit{lamp}$ of~\cref{fig:timedlamp}, which is model-checked against the following formulae (over finite or infinite timed words):
\begin{itemize}
\item $\alpha_1 = \globally (\textit{push} \Rightarrow \eventually_{\geq 3} \textit{blink})$: Each time the button is pushed, regardless of whether it is pushed again later, the lamp must blink after $3$ time units. This clearly holds in $\mathcal{M}_\textit{lamp}$.

\item $\alpha_2 = \globally (\textit{push} \land \globally_{(0, \infty)}(\neg \textit{push}) \Rightarrow \eventually_{= 3} \textit{blink})$:
Each time the button is pushed, if it is not pushed again strictly later, the lamp must blink after exactly $3$ time units.
Like $\alpha_1$, this clearly holds in $\mathcal{M}_\textit{lamp}$.

\item $\alpha_3 = \globally (\textit{push} \land \globally_{(0, \infty)}(\neg \textit{push}) \Rightarrow \neg (\eventually_{(2, 3)} \textit{blink}))$:
Same premise as in $\alpha_2$, but the lamp must \emph{not} blink in $(t + 2, t + 3)$.
This holds in $\mathcal{M}_\textit{lamp}$ as it is clear that the lamp will only blink at $t+1, t+2, t+3$.

\item $\alpha_4 = \globally (\textit{push} \land (\nextx_{(0, \infty)} \textit{push}) \land \eventually_{[0, 1)} (\textit{push} \land \globally_{(0, \infty)}(\neg \textit{push})) \Rightarrow \neg (\eventually_{= 3} \textit{blink}))$:
Each time the button is pushed at time $t$, if it is pushed for the last time at some $t' \in (t, t + 1)$, the lamp must \emph{not} blink at $t + 3$.
This holds in $\mathcal{M}_\textit{lamp}$ as it is clear that the lamp will only blink at $t'+1, t'+2, t'+3$.

\item $\alpha_5 = \globally (\textit{push} \land (\nextx_{(0, \infty)} \textit{push}) \land \eventually_{[0, 1)} (\textit{push} \land \globally_{(0, \infty)}(\neg \textit{push})) \Rightarrow \neg (\eventually_{[4, 5]} \textit{blink}))$:
Same premise as in $\alpha_4$, but the lamp must \emph{not} blink in $[t + 4, t + 5]$.
This holds in $\mathcal{M}_\textit{lamp}$ as $t' + 3 < t + 4$.

\item $\alpha_6 = \globally(\textit{blink} \land \globally_{(0, \infty)} \textit{blink} \Rightarrow \once_{=3} (\textit{push}))$: 
If the lamp blinks for the last time at $t$, the button must have been pushed at $t - 3$.
This holds in $\mathcal{M}_\textit{lamp}$.

\item $\alpha_7 = \globally(\textit{blink} \land \globally_{(0, \infty)} \textit{blink} \Rightarrow \neg \once_{=2} (\textit{push}))$: 
Same premise as in $\alpha_6$, but the button must \emph{not} have been pushed at $t - 2$.
This holds in $\mathcal{M}_\textit{lamp}$ since if the button was pushed at $t - 2$,
the last blink must happen at $\geq t + 1$.

\item $\alpha_8 = \globally(\textit{push} \land \textit{blink} \Rightarrow \eventually_{\leq 5} (\neg \textit{blink} \land \globally_{[0, 2]} (\neg \textit{blink})))$: Each time the button is pushed, the lamp goes off (i.e.~$\neg \textit{blink}$ occurs) for $2$ time units at least once in the next $5$ time units. This does not hold in $\mathcal{M}_\textit{lamp}$ since
$(\{\textit{push}, \textit{blink}\}, 0)(\{\textit{blink}\}, 1)(\{\textit{blink}\}, 2)(\{\textit{blink}\}, 3)$ is a valid behaviour
in the finite-word case, and the user can push the button repeatedly every, e.g.,~$3.9$ time units in the infinite-word case.

\item $\alpha_9 = \globally(\textit{blink} \land \eventually_{\leq 1} \textit{blink} \land \globally_{\leq 9} (\eventually_{\leq 1} \textit{blink}) \Rightarrow \PnF_{\leq 9} ( \textit{push},\textit{push}))$:
    If the lamp blinks continuously for $9$ time units, then the button must have been pressed at least twice in this period.
    This holds in $\mathcal{M}_\textit{lamp}$.
    
\item $\alpha_{10} = \globally(\textit{blink} \land \eventually_{\leq 1} \textit{blink} \land \globally_{\leq 9} (\eventually_{\leq 1} \textit{blink}) \Rightarrow \PnF_{\leq 9} ( \textit{push},\textit{push},\textit{push}))$:
    This does not hold as one may, e.g.,~push the button every $3.9$ time units.
\end{itemize}    
The results are in~\cref{tab:timedlamp}. 
$\mightyppl{}$ finishes model checking each of the $10$ properties in $<0.4$s in every configuration; \textsc{Tempora}
supports $4$ of them but performs worse on $\alpha_3, \alpha_5$, and produces a wrong answer on $\alpha_8$.

\begin{figure}[!htbp]
\centering
\begin{Verbatim}[fontsize=\scriptsize]
# M
process:M
clock:1:x
clock:1:y
location:M:ell_0{initial: : labels: accept_M}
location:M:ell_1{}
edge:M:ell_0:ell_0:a{provided: p_push %
edge:M:ell_0:ell_1:a{provided: p_push >= 1 && p_blink >= 1 : do: x = 0; y = 0}
edge:M:ell_1:ell_1:a{provided: p_push %
edge:M:ell_1:ell_1:a{provided: p_push >= 1 && p_blink >= 1 : do: x = 0; y = 0}
edge:M:ell_1:ell_0:a{provided: p_push %
\end{Verbatim}
\captionof{listing}{$\mathcal{M}_\textit{lamp}$ in \textsc{TChecker} format.}
\end{figure}

\begin{table}[!htbp]
\captionof{table}{Execution times on the timed lamp benchmarks.
Times are in seconds.
`TO' indicates timeouts (600s).
`-' means unsupported.
Incorrect results are marked by `*'.}
\label{tab:timedlamp}
\centering
\scalebox{0.8}{
\def\arraystretch{1.0}
\setlength\tabcolsep{1mm}
\begin{tabular}{l@{\hspace{2mm}}r>{\columncolor{mySkyBlue!10}}r>{\columncolor{myYellow!10}}r>{\columncolor{myBlue!45}}r>{\columncolor{myYellow!45}}r>{\columncolor{myBlue!45}}r>{\columncolor{myYellow!45}}r}
\toprule %
&
 & \cellcolor{white}\multirowcell{3}{$\tempora{}$ \\ \small \texttt{infinite}\\{}}
& \cellcolor{white}\multirowcell{3}{$\tempora{}$ \\ \small \texttt{finite}\\{}}
 & \cellcolor{white}\multirowcell{3}{$\mightyppl{}$ \\ \small \texttt{tck -{}-inf} \\ \small \texttt{compflat}}
 & \cellcolor{white}\multirowcell{3}{$\mightyppl{}$ \\ \small \texttt{tck -{}-fin} \\ \small \texttt{compflat}} 
 & \cellcolor{white}\multirowcell{3}{$\mightyppl{}$ \\ \small \texttt{tck -{}-inf} \\ \small \texttt{flat}}
 & \cellcolor{white}\multirowcell{3}{$\mightyppl{}$ \\ \small \texttt{tck -{}-fin} \\ \small \texttt{flat}} \\ \\
$\varphi$  & Hold?   \\
\otoprule
   $\alpha_1$       & \cmark         & 0.002     & 0.000    & 0.009      & 0.009        & 0.014        & 0.009    \\
  $\alpha_2$       & \cmark          & -         & -        & 0.009      & 0.010        & 0.033        & 0.016    \\
  $\alpha_3$       & \cmark          & 6.834     & 0.037    & 0.008      & 0.009        & 0.027        & 0.014           \\
  $\alpha_4$     & \cmark            & -         & -        & 0.013      & 0.013        & 0.055        & 0.023    \\
   $\alpha_5$       & \cmark         & TO        & 2.572    & 0.013      & 0.014        & 0.055        & 0.022       \\
  $\alpha_6$       & \cmark          & -         & -        & 0.027      & 0.016        & 0.061        & 0.025    \\
  $\alpha_7$       & \cmark          & -         & -        & 0.012      & 0.011        & 0.027        & 0.014      \\
  $\alpha_8$     & \xmark            & 0.003     & 0.000*   & 0.009      & 0.011        & 0.027        & 0.013     \\
   $\alpha_9$       & \cmark         & -         & -        & 0.023      & 0.022        & 0.259        & 0.079   \\
  $\alpha_{10}$       & \xmark       & -         & -        & 0.014      & 0.031        & 0.370        & 0.109    \\
\bottomrule
\end{tabular}
}
\end{table}

\paragraph{Fischer's mutual exclusion protocol.}

This is a classic mutual exclusion algorithm (from~\citep{lamport1987fast})
that has been used as a standard example for $\ta{}$-based
verification tools such as \textsc{Uppaal}.
We use the file \texttt{fischer.sh} from~\citep{TChecker} to generate the system models (asynchronous networks of $\ta{s}$ : one $\ta{}$ $\mathcal{P}_i$ for each of the $N$ processes) with a fixed delay $K = 5$; then we add the atomic propositions $\textit{idle}_i, \textit{req}_i, \textit{wait}_i, \textit{cs}_i$ to indicate which location the process $\mathcal{P}_i$ is in. For example, $\textit{cs}_i$ holds on the transition from $\ell_\textit{req}$ to $\ell_\textit{cs}$ in  $\mathcal{P}_i$. The system models are model-checked against the following formulae (over infinite timed words):
\begin{itemize}

    \item $\beta_1 = \bigwedge_{i} \big( \globally(\textit{req}_i \Rightarrow \eventually_{[0,10]} (\textit{cs}_i \land \nm (\neg \textit{cs}_i))) \big)$:  
    After $\mathcal{P}_i$ issues a request, it must enter the critical section in $10$ time units. This does not hold as a process can wait in $\ell_\textit{wait}$ for arbitrarily long before entering $\ell_\textit{cs}$.
    
    \item $\beta_2 = \bigwedge_{i} \big( \globally(\textit{cs}_i \land \nm (\neg \textit{cs}_i) \Rightarrow \past_{[5,10]} \textit{req}_i) \big)$:  
   Whenever $\mathcal{P}_i$ enters the critical section, it must have issued a request between $5$ and $10$ time units earlier. This does not hold for the same reason as $\beta_1$.
    
    \item $\beta_3 = \bigwedge_{i} \big(\globally(\textit{cs}_i \land \nm (\neg \textit{cs}_i) \Rightarrow \neg \PnF_{[0,11]} (\textit{cs}_i \land \nm (\neg \textit{cs}_i), \textit{cs}_i \land \nm (\neg \textit{cs}_i)))\big)$:  
    No process $i$ can enter the critical section more than three times within any $11$ time units window. This does not hold as a process can delay for $0$ time units in $\ell_\textit{idle}$ and $\ell_\textit{req}$, for exactly $5.1$ time units in $\ell_\textit{wait}$, and repeat this pattern. 
    
    \item $\beta_4 = \bigwedge_{i} \big(\globally(\textit{cs}_i \land \nm (\neg \textit{cs}_i) \Rightarrow \neg \PnF_{[0,10]} (\textit{cs}_i \land \nm (\neg \textit{cs}_i), \textit{cs}_i \land \nm (\neg \textit{cs}_i)))\big)$:  
    Same as $\alpha'_3$ but with $10$ time units windows. This holds as a process must take $> 5$ time units from $\ell_\textit{wait}$ to $\ell_\textit{cs}$.
    
    \item $\beta_5 = \bigwedge_{i} \big(\globally(\textit{cs}_i \land \nm (\neg \textit{cs}_i) \Rightarrow
    \once(\textit{wait}_i \land \once_{[0, 5]} (\textit{req}_i \land \neg \nm (\textit{wait}_i))))\big)$:  
   Whenever $\mathcal{P}_i$ enters the critical section, it must have issued an \emph{initial} request, which is either just the first event or preceded by $\textit{idle}_i$. This holds.
    
    \item $\beta_6 = \bigwedge_{i} \big(\globally(\textit{cs}_i \land \nm (\neg \textit{cs}_i) \Rightarrow
    \once_{[5, 20]} (\textit{wait}_i \land \once_{[0, 5]} (\textit{req}_i \land \neg \nm (\textit{wait}_i))))\big)$:  
    Same as $\beta_5$ but the initial request must be between $5$ and $25$ time units earlier. This does not hold for the same reason as $\beta_1$.
\end{itemize}
The results are in~\cref{tab:fischer}.
While \textsc{Tempora} demonstrates noticeable performance gains on $\beta_5$, it produces incorrect results for the closely related $\beta_6$.

\begin{figure}[!htbp]
\centering
\begin{Verbatim}[fontsize=\scriptsize]
# Process 1
process:P1
clock:1:x1
location:P1:idle{initial:}
location:P1:req{invariant:x1<=5}
location:P1:wait{}
location:P1:cs{labels:cs1}
edge:P1:idle:req:a{provided: id==0 : do: x1=0; p_req_1 = 1; p_wait_1 = 0; p_cs_1 = 0; p_idle_1 = 0}
edge:P1:req:wait:a{provided: x1<=5 : do: x1=0; id=1; p_req_1 = 0; p_wait_1 = 1; p_cs_1 = 0; p_idle_1 = 0}
edge:P1:wait:req:a{provided: id==0 : do: x1=0; p_req_1 = 1; p_wait_1 = 0; p_cs_1 = 0; p_idle_1 = 0}
edge:P1:wait:cs:a{provided: x1>5 && id==1 : do: p_req_1 = 0; p_wait_1 = 0; p_cs_1 = 1; p_idle_1 = 0}
edge:P1:cs:idle:a{do: id=0; p_req_1 = 0; p_wait_1 = 0; p_cs_1 = 0; p_idle_1 = 1}

# Process 2
process:P2
clock:1:x2
location:P2:idle{initial:}
location:P2:req{invariant:x2<=5}
location:P2:wait{}
location:P2:cs{labels:cs2}
edge:P2:idle:req:a{provided: id==0 : do: x2=0; p_req_2 = 1; p_wait_2 = 0; p_cs_2 = 0; p_idle_2 = 0}
edge:P2:req:wait:a{provided: x2<=5 : do: x2=0; id=2; p_req_2 = 0; p_wait_2 = 1; p_cs_2 = 0; p_idle_2 = 0}
edge:P2:wait:req:a{provided: id==0 : do: x2=0; p_req_2 = 1; p_wait_2 = 0; p_cs_2 = 0; p_idle_2 = 0}
edge:P2:wait:cs:a{provided: x2>5 && id==2 : do: p_req_2 = 0; p_wait_2 = 0; p_cs_2 = 1; p_idle_2 = 0}
edge:P2:cs:idle:a{do: id=0; p_req_2 = 0; p_wait_2 = 0; p_cs_2 = 0; p_idle_2 = 1}
\end{Verbatim}
\captionof{listing}{$\mathcal{P}_1, \mathcal{P}_2$ in \textsc{TChecker} format.}
\end{figure}

\begin{table}[!htbp]                    
\captionof{table}{Execution times on the Fischer benchmarks.
Times are in seconds.
`TO' indicates timeouts (600s).
`-' means unsupported.
Incorrect results are marked by `*'.
}%
\label{tab:fischer}
\centering
\scalebox{0.8}{
\def\arraystretch{1.0}
\setlength\tabcolsep{2mm}
\begin{tabular}{l@{\hspace{2mm}}rr>{\columncolor{mySkyBlue!10}}r>{\columncolor{myBlue!45}}r}
\toprule 
& & & \cellcolor{white}\multirowcell{3}{$\tempora{}$ \\ \small \texttt{infinite}\\{}}
& \cellcolor{white}\multirowcell{3}{$\mightyppl{}$ \\ \small \texttt{tck -{}-inf} \\ \small \texttt{compflat}} \\ \\
$\varphi$  & \cellcolor{white}$N$ & \cellcolor{white}Hold? \\
\otoprule
$\beta_1$  & 2     &  \xmark      & 0.141    & 0.010         \\ 
$\beta_1$  & 3     &  \xmark      & 484.967  & 0.029        \\ 
$\beta_1$  & 4     &  \xmark      & TO       & 0.087         \\ 
$\beta_2$  & 2     &  \xmark      & 0.062*   & 0.266           \\ 
$\beta_2$  & 3     &  \xmark      & 112.774* & TO         \\ 
$\beta_2$  & 4     &  \xmark      & TO       & TO        \\ 
$\beta_3$  & 2     &  \xmark      & -        & 0.156         \\ 
$\beta_3$  & 3     &  \xmark      & -        & 6.504         \\ 
$\beta_3$  & 4     &  \xmark      & -        & TO           \\ 
$\beta_4$  & 2     &  \cmark      & -        & 0.577         \\ 
$\beta_4$  & 3     &  \cmark      & -        & 161.910               \\ 
$\beta_4$  & 4     &  \cmark      & -        & TO        \\ 
$\beta_5$  & 2     &  \cmark      & 0.056    & 3.841         \\ 
$\beta_5$  & 3     &  \cmark      & 33.281   & TO            \\ 
$\beta_5$  & 4     &  \cmark      & TO       & TO            \\ 
$\beta_6$  & 2     &  \xmark      & 0.031*   & 2.957         \\ 
$\beta_6$  & 3     &  \xmark      & 8.136*   & TO             \\ 
$\beta_6$  & 4     &  \xmark      & TO       & TO              \\ 
\bottomrule
\end{tabular}
}
\end{table} 

\paragraph{Pinwheel scheduling.}

 The \emph{pinwheel scheduling} problem~\citep{Holte1989pinwheel} asks
 whether there is an \emph{infinite} schedule for tasks $\{i \mid 1 \leq i \leq k\}$ (each with a \emph{relative deadline} $a_i \geq 2$),
e.g.,~on each day we can schedule a task, and
 each task is scheduled at least once in every $a_i$ days.
 We use the model $\ta{}$ $\mathcal{M}_\textit{pin}^k$ to enforce that exactly one of $\{p_1, \dots, p_k\}$ is scheduled in each slot, and any two scheduled tasks must be separated by at least $1$ time unit. We then model-check $\mathcal{M}_\textit{pin}^k$ against the corresponding formula
 for an instance $(a_1, \dots, a_k)$ of the problem:
 \[
\varphi_{(a_1, \dots, a_k)}  = (\bigwedge_{i \in \{1, \dots, k\}} \eventually p_i) 
    \Rightarrow \neg \bigwedge_{i \in \{1, \dots, k\}} \globally (p_i \Rightarrow \nextx \eventually_{[0, a_i]} p_i) \;.
\]
The results are in~\cref{tab:pinwheel}.
To highlight the effect of using multiple cores, here we 
omit the comparison with $\textsc{Tempora}$ (which times out on all the instances tested) and 
exclude the time used for flattening and compiling (which on the most difficult instances take $\sim$ $110$s). The single-threaded performance of \textsc{TChecker} in the `\texttt{flat}' mode is roughly comparable to that of \textsc{LTSmin}.
 The latter, however, can run on multiple threads and we can see that a speedup is generally observed as the number of threads increases, and this improvement is sometimes linear with respect to the thread count.  Furthermore, we notice that using more than $16$ threads does not always yield significant performance gains, likely due to hardware limitations and increased synchronisation overhead.

 \begin{table}[!htbp]
 \caption{Execution times on the pinwheel scheduling benchmarks. 
 Times are in seconds. Numbers in the heading are the numbers of threads. `TO' indicates timeouts (600s).}
 \label{tab:pinwheel}
 \centering
 \scalebox{0.76}{
 \def\arraystretch{1.0}
 \setlength\tabcolsep{1mm}
 \begin{tabular}{l@{\hspace{2mm}}r>{\columncolor{myBlue!45}}r>{\columncolor{myBlue!45}}r>{\columncolor{myBlue!45}}r>{\columncolor{myBlue!45}}r>{\columncolor{myBlue!45}}r>{\columncolor{myBlue!45}}r>{\columncolor{myBlue!45}}r>{\columncolor{myBlue!45}}r}
 \toprule %
 &
 & \cellcolor{white}\multirowcell{3}{$\mightyppl{}$ \\ \small \texttt{tck -{}-inf} \\ \small \texttt{compflat}}
 & \cellcolor{white}\multirowcell{3}{$\mightyppl{}$ \\ \small \texttt{tck -{}-inf} \\ \small \texttt{flat}}
 & \multicolumn{6}{c}{\cellcolor{white}\multirowcell{3}{$\mightyppl{}$ \\ \small \texttt{opaal2lts-mc} \\ \small \texttt{flat}}}  \\ \\ \\
 \cmidrule[\heavyrulewidth](l{2mm}r{2mm}){5-10}
 $\varphi$ & Hold? & \cellcolor{white} & \cellcolor{white} & \cellcolor{white}1 & \cellcolor{white}2 & \cellcolor{white}4 & \cellcolor{white}8 & \cellcolor{white}16 & \cellcolor{white}32 \\
 \otoprule
 $(3, 4, 5, 7)$             & \cmark       & 1.899    & 0.068      & 0.028   & 0.027   & 0.036   & 0.063     & 0.145   & 0.428       \\
 $(3, 4, 5, 8)$             & \xmark       & 0.291    & 0.059      & 0.013   & 0.033   & 0.036   & 0.063     & 0.145   & 0.305          \\
 $(2, 7, 9, 9, 16)$         & \cmark       & TO       & 1.087      & 0.912   & 0.519   & 0.319   & 0.286     & 0.452   & 0.748             \\
 $(2, 7, 9, 10, 16)$        & \xmark       & 11.154   & 0.331      & 0.063   & 0.043   & 0.050   & 0.075     & 0.156   & 0.333  \\
 $(3, 7, 8, 8, 9, 10)$      & \cmark       & TO       & 10.881     & 9.106   & 4.789   & 2.814   & 2.122     & 3.254   & 3.890  \\
 $(3, 7, 8, 8, 9, 11)$      & \xmark       & TO       & 2.484      & 0.474   & 0.356   & 0.242   & 0.232     & 0.336   & 0.637  \\
 $(4, 6, 8, 9, 9, 9, 15)$   & \cmark       & TO       & 202.238    & TO      & TO      & 375.863 & 312.715   & 392.217 & 475.140  \\
 $(4, 6, 8, 9, 9, 9, 16)$   & \xmark       & TO       & 16.072     & 4.467   & 3.426   & 0.269   & 0.486     & 1.971   & 1.937       \\
 $(5, 5, 8, 9, 9, 9, 11)$   & \cmark       & TO       & 125.591    & 491.325 & 260.836 & 153.255 & 120.500   & 174.310 & 201.452      \\
 $(5, 5, 8, 9, 9, 9, 12)$   & \xmark       & TO       & 15.704     & 3.014   & 1.121   & 1.799   & 0.480     & 3.263   & 9.223     \\
 $(6, 7, 7, 7, 8, 8, 10)$   & \cmark       & TO       & 78.520     & 210.064 & 113.902 & 64.913  & 48.763    & 70.275  & 77.565    \\
 $(6, 7, 7, 7, 8, 8, 11)$   & \xmark       & TO       & 16.579     & 8.268   & 2.543   & 2.471   & 2.895     & 5.219   & 3.431        \\
 \bottomrule
 \end{tabular}
 }
 \end{table}

\subsubsection*{Acknowledgements.}
The work of Khushraj Madnani was supported by the ANRF ARG-MATRICS Grant (File No.
ANRF/ARGM/2025/003152/MTR) under the project titled ``Approaching the Decidability Purlieu of Timed
Languages (ADePT).''

\subsubsection*{Data availability statement.}
The tool implementation, benchmark models, and datasets generated during the current study are available in the Zenodo repository, \url{https://doi.org/10.5281/zenodo.19852060}.

\newpage
\renewcommand{\bibsection}{\section*{References}} %
\bibliographystyle{splncs04nat}
\begingroup
  \microtypecontext{expansion=sloppy}
  \small %
  \bibliography{biblio}
\endgroup

\end{document}